\documentclass[a4paper]{article}

\usepackage{amssymb}
\usepackage{amsfonts}
\usepackage{cite}
\usepackage{epsfig}

\newcommand{\be}{\begin{equation}}
\newcommand{\ee}{\end{equation}}
\newcommand{\fig}[4]{\begin{figure}[ht]\epsfxsize=#2\bigskip\centerline{\epsfbox{#1}}\caption{\small\it #3 \label{#4}}\bigskip\end{figure}}
\newcommand{\ba}[1]{\begin{array}{#1}}
\newcommand{\ea}{\end{array}}

\begin{document}

\begin{center}

{\ \\[5mm]\Large\bf\sc\noindent An Expanding Locally Anisotropic (ELA) Metric Describing Matter in an Expanding Universe\\[20mm]}

{P. Castelo Ferreira\footnote{pedro.castelo.ferreira@ist.utl.pt\\ \indent\ \  CENTRA--IST, Av. Rovisco Pais 1, 1100 Lisboa, Portugal\\ \indent\ \  Eng. Electr. -- ULHT, Campo Grande 376, 1749-024 Lisboa, Portugal}\\[30mm]}

{\bf\sc Abstract}
\end{center}
\noindent It is suggested an expanding locally anisotropic metric (ELA) ansatz describing matter in a flat expanding
universe which interpolates between the Schwarzschild (SC) metric near
point-like central bodies of mass $M$ and the Robertson-Walker (RW) metric
for large radial coordinate:\\[-4mm]
\begin{center}$
\displaystyle ds^2=Z\,c^2dt^2-\frac{1}{Z}\left(dr_1-\frac{H\,r_1}{c}\,Z^{\frac{\alpha}{2}+\frac{1}{2}}\,cdt\right)^2-r_1^2\,d\Omega\ ,
$\end{center}
where $Z=1-U_{\mathrm{SC}}$ with $U_{\mathrm{SC}}=2GM/(c^2r_1)$, $G$ is the Newton constant,
$c$ is the speed of light, $H=H(t)=\dot{a}/a$ is the time-dependent Hubble rate,
$d\Omega=d\theta^2+\sin^2\theta\,d\varphi^2$ is the solid angle element, $a$ is the universe scale factor
and we are employing the coordinates $r_1=ar$, being $r$ the radial coordinate for which the
RW metric is diagonal. For constant exponent $\alpha=\alpha_0=0$ it is retrieved the
isotropic McVittie (McV) metric and for $\alpha=\alpha_0=1$ it is retrieved the locally anisotropic
Cosmological-Schwarzschild (SCS) metric, both already discussed in the literature. However it is shown
that only for constant exponent $\alpha=\alpha_0> 1$ exists an event horizon at the SC radius $r_1=2GM/c^2$ and only for $\alpha=\alpha_0\ge 3$ space-time is singularity free for this value of the radius. These bounds exclude the previous existing metrics, for which the SC radius is a naked extended singularity. In addition it is shown that for $\alpha=\alpha_0>5$ space-time is approximately Ricci flat in a neighborhood of the event horizon such that the SC metric is a good approximation in this neighborhood. It is further
shown that to strictly maintain the SC mass pole at the origin $r_1=0$ without
the presence of more severe singularities it is required a radial coordinate dependent correction
to the exponent $\alpha(r_1)=\alpha_0+\alpha_1\,2GM/(c^2\,r_1)$ with a negative coefficient $\alpha_1<0$.
The energy-momentum density, pressures and equation of state are discussed.

\thispagestyle{empty}
\newpage

\section{Introduction}
Following the discovery by Hubble of the universe expansion~[\citenum{Hubble}] several global metrics
describing cosmological expansion were derived, namely for flat~[\citenum{RW}], closed
and open~[\citenum{Lemaitre}] isotropic universes. These results raised the question whether local astrophysical
systems are affected by the global universe expansion. In order to investigate this issue it
is required to consider a metric that describes both local matter distributions, namely point-like central
masses, and global expansion, hence interpolating between the Schwarzschild metric~[\citenum{Schwarzschild}]
(SC metric) near a central mass $M$ and the specific metric describing global expansion.
One of the first studies dealing with this problem was carried by McVittie
assuming global isotropy~[\citenum{McVittie}] and employing isotropic coordinates. However the McV
metric has naked space-time singularities at the SC radius~[\citenum{sing}] such that space-time is not
complete and the mass inside a shell of finite radius is divergent. More recent developments on
this problem considered the locally anisotropic CSC metric~[\citenum{cosm_SC}]. Experimentally there
is today evidence for local anisotropy, both from direct astrophysical measurements~[\citenum{anisotropy}] as well as from the measurements of the cosmological microwave background
radiation~[\citenum{wmap}], hence local isotropy is not mandatory as long as global isotropy
is maintained. However the SCS metric has the same singularity behavior of the McV metric.
In the following we will analyze this problem obtaining an Expanding Locally Anisotropic (ELA) metric
describing a singularity free space-time (except for the SC mass-pole) which has, as particular cases,
both the isotropic McV metric and the locally anisotropic SCS metric.

According to today's most recent data the universe is flat~[\citenum{wmap}], hence global expansion is described by the RW
metric~[\citenum{RW}] (RW metric). We will work with radial coordinate $r_1=ar$, being $r$ the coordinate
for which the RW metric is diagonal. The radial coordinate $r_1$ directly correspond to the measurable
distances (it is an area radius). Following this convention the SC metric line-element is 
\be
ds^2=c^2\,Z\,dt^2-\frac{1}{Z}\,dr_1^2-r_1^2\,d\Omega\ ,
\label{SC_metric}
\ee
and the RW metric line-element describing a flat expanding universe is
\be
ds^2=c^2\,(1-N_{\mathrm{RW}}^2)\,dt^2+2c\,N_{\mathrm{RW}}\,dr_1\,dt-dr_1^2-r_1^2\,d\Omega\ ,
\label{RW_metric}
\ee
where $Z=1-U_{SC}$, $U_{\mathrm{SC}}$ is the Schwarzschild gravitational potential and
$N_{\mathrm{RW}}$ is the Robertson-Walker shift function
\be
U_{\mathrm{SC}}=\frac{2GM}{c^2r_1}\ \ ,\ \ N_{\mathrm{RW}}=H\,\frac{r_1}{c}\ .
\label{Nshift}
\ee
$G$ is the Newton constant, $c$ is the speed of light, $H=\dot{a}/a$ is the time-varying Hubble
rate defined as the rate of change of the universe scale factor $a$ and $d\Omega=d\theta^2+\sin^2\theta\,d\varphi^2$
is the solid angle line-element. The only singularity of the space-time described by the SC
metric~(\ref{SC_metric}) is the mass-pole (of value $M$) at the origin, the scalar curvature
$R=g^{\alpha\beta}R_{\alpha\beta}$ (the Ricci scalar) and the curvature invariant
${\mathcal{R}}=R^{\mu\nu\lambda\rho}R_{\mu\nu\lambda\rho}$ are
\be
R_{\mathrm{SC}}=-\frac{8\pi\,G}{c^2}\,M\,\delta^{(3)}(r_1)\ \ ,\ \ {\mathcal{R}}_{\mathrm{SC}}=\frac{48 (GM)^2}{c^2\,r_1^6}\ .
\label{SC_R}
\ee
As for the space-time described by the RW metric~(\ref{RW_metric}) is singularity free, the scalar curvature and curvature invariant are time-dependent only, $R_{\mathrm{RW}}=-6(1-q)H^2/c^2$ and ${\mathcal{R}}_{\mathrm{RW}}=12(1-q^2)H^4/c^4$, where $q=-(\dot{H}/H^2+1)=-a\ddot{a}/\dot{a}^2$ is the time varying universe deceleration factor~[\citenum{Sandage}]. The SC metric has an event horizon at the SC radius $r_1=r_{1.\mathrm{SC}}=2GM/c^2$ and the RW metric has an event horizon at the Hubble radius $r_1=l_H=c/H$.

The McV metric~[\citenum{McVittie}] and the SCS metric~[\citenum{cosm_SC}] line-elements are, respectively,
\be
\ba{rcl}
ds_{\mathrm{McV}}^2&=&\displaystyle c^2\,Z\,dt^2-\frac{1}{Z}\left(dr_1-c\,N_{\mathrm{RW}}\,Z^\frac{1}{2}\,dt\right)^2-r_1^2\,d\Omega\ ,\\[3mm]
ds_{\mathrm{SCS}}^2&=&\displaystyle c^2\,Z\,dt^2-\frac{1}{Z}\left(dr_1-c\,N_{\mathrm{RW}}\,Z\,dt\right)^2-r_1^2\,d\Omega\ .
\ea
\label{McV_SCS_metric}
\ee
The singular behavior of these two metrics at the origin $r_1=0$ and at the SC radius $r_1=r_{1.\mathrm{SC}}$ is obtained from the curvature invariant ${\mathcal{R}}$.
The leading expressions for these quantities are
\be
\ba{lcl}
\displaystyle R_{\mathrm{McV}}(\sim 0)\sim-\frac{8\pi\,G}{c^2}\,M\,\delta^{(3)}(r_1)&,& \displaystyle {\mathcal{R}}_{\mathrm{McV}}(\sim r_{1.\mathrm{SC}})\sim\frac{12(1-q)^2H^4}{Z}\ ,\\
\displaystyle R_{\mathrm{SCS}}(\sim 0)\sim-\frac{8\pi\,G}{c^2}\,M\,\delta^{(3)}(r_1)&,& \displaystyle{\mathcal{R}}_{\mathrm{SCS}}(\sim r_{1.\mathrm{SC}})\sim\frac{4(1+q)^2\,H^4 (GM)^2}{c^4\,r_1^2\,Z^2}\ ,
\label{McV_SCS_R}
\ea
\ee
At the origin the singularities coincide with the Schwarzschild mass-pole, ${\mathcal{R}}_{\mathrm{McV}}(r_1\to 0)={\mathcal{R}}_{\mathrm{SCS}}(r_1\to 0)= {\mathcal{R}}_{\mathrm{SC}}$~(\ref{SC_R}) and these space-times are singularity free outside the SC radius up to infinity, $r_1>r_{1.\mathrm{SC}}$, the curvature invariants are finite. However, at the SC radius $r_1=r_{1.\mathrm{SC}}$, the space-time described by these two metrics have extended singularities
(over a sphere of fixed radius) such that space-time is not complete and the mass inside of a shell of
finite radius $r_1>r_{1.\mathrm{SC}}$ is divergent. Moreover these are naked singularities, the non-null connections $\Gamma^1_{00}$ and $\Gamma^1_{01}$ do not vanish at $r_1=r_{1.\mathrm{SC}}$. For the McV and
SCS metric we obtain, respectively,
\be
\ba{lcl}
\displaystyle\,_{(\mathrm{McV})}\Gamma^1_{00}( r_{1.\mathrm{SC}})=-r_{1.\mathrm{SC}}\frac{H^2}{2c^2}+r_{1.\mathrm{SC}}^3\frac{H^4}{c^4}&,&\displaystyle \,_{(\mathrm{McV})}\Gamma^1_{01}(r_{1.\mathrm{SC}})=\infty\\
\displaystyle\,_{(\mathrm{SCS})}\Gamma^1_{00}(r_{1.\mathrm{SC}})=0&,&\displaystyle\,_{(\mathrm{SCS})}\Gamma^1_{01}(r_{1.\mathrm{SC}})=2H/c.
\ea
\label{connections}
\ee

Expansion effects are negligible for relatively short scales astrophysical systems, the gravitational attraction law for space-times described by the SC metric, describes to a very good accuracy the dynamics of most astrophysical systems. Hence a singularity near massive objects is unwelcome, it means that the
expansion effects are dominant at the SC radius. From a theoretical perspective, naked extended
singularities and divergent masses are also unwelcome and do not correspond to any physical meaningful
reality. We will next suggest a metric ansatz for which the event horizon is maintained at the SC radius
and space-time is singularity free except for the mass-pole at the origin.

\section{The Metric Ansatz}
We note that both the McV metric and the SCS metric~(\ref{McV_SCS_metric}) discussed above,
are obtained from the SC metric~(\ref{SC_metric}) by considering a shift function which is the product of the RW shift function $N_{\mathrm{RW}}$ by a power of the factor $Z=1-U_{\mathrm{SC}}$.
This structure of the shift function ensures that:
\begin{enumerate}
\item both spherical symmetry and the coordinate measure $\sqrt{-g}=r_1^2\sin\theta$ are maintained;
\item at spatial infinity the metric converges asymptotically to the RW metric~(\ref{RW_metric}), $\displaystyle\lim_{r_1\to\infty}Z= 1$;
\item for positive exponent $\alpha>0$ of the factor $Z$ at the SC radius, the metric converges to the SC metric~(\ref{SC_metric}), $\displaystyle\lim_{r_1\to r_{1.\mathrm{SC}}}Z= 0$.
\end{enumerate}
Hence, a straight forward generalization of metrics~(\ref{McV_SCS_metric}) is
\be
ds^2=c^2\,Z\,dt^2-\frac{1}{Z}\left(dr_1-c\,N_{\mathrm{RW}}\,Z^{\frac{\alpha}{2}+\frac{1}{2}}\,dt\right)^2-r_1^2\,d\Omega\ ,
\label{ELA_metric}
\ee
where $\alpha$ is an arbitrary exponent. The McV metric is obtained for constant exponent
$\alpha=\alpha_0=0$ and the CSC metric for $\alpha=\alpha_0=1$.

We proceed to analyze the space-time represented by this metric. First we consider constant exponent $\alpha=\alpha_0$. We remark that the SC event horizon is maintained as long as the
connections $\,_{(\alpha_0)}\Gamma^1_{\ 0\mu}$ vanish at the SC radius, $r_1=r_{1.\mathrm{SC}}$. These conditions are obeyed for the lower bound $\alpha_0>1$, for which the SC radius is an event horizon. As already discussed this result excludes both the McV and SCS metrics~(\ref{connections}).

The space-time singularity behavior is obtained as usual from the curvature
invariant ${\mathcal{R}}$. At the Schwarzschild radius
$r_1=r_{1.\mathrm{SC}}$, the limiting expressions for this quantity are
\be
\ba{l}
\displaystyle{\mathcal{R}}_{\alpha_0}(r_1\to r_{1.\mathrm{SC}})\sim\\
\ \sim\left\{
\ba{lcl}
0&,&\alpha_0\in]3,+\infty[\\
12U_{\mathrm{SC}}^{-4}-12U_{\mathrm{SC}}^{-2}(1+q)H^2+9(1+q)^2H^4&,&\alpha_0=3\\
Z^{\alpha_0-3}\ \ \ \to\infty&,&\alpha_0\in[1,3[\\
Z^{2\alpha_0-4}\ \ \to\infty&,&\alpha_0\in]-\infty,1[/\{0\}\\
Z^{-1} \ \ \ \ \ \, \to\infty&,&\alpha_0=0\\
\ea\right.
\ea
\ee
Only for $\alpha_0\ge 3$ the curvature invariant is finite, hence we conclude that the
event horizon is singularity free only for this condition. In addition for $\alpha_0>5$ both
the scalar curvature and its derivative, as well all the components of the Ricci tensor, vanish at $r_1=r_{1.\mathrm{SC}}$, $R_{(\alpha_0>5)}(r_{1.\mathrm{SC}})=R'_{(\alpha_0>5)}(r_{1.\mathrm{SC}}) = \,_{(\alpha_0>5)}R_{\mu\nu}(r_{1.\mathrm{SC}})=0$ (here prime denotes differentiation with respect to $r_1$) such that space-time is approximately Ricci flat in a neighborhood of the event horizon, hence coinciding with the properties of the space-time described by the SC metric\footnote{We recall that the background of the SC metric corresponds to empty Ricci flat space-time, hence with $R=R'=R_{\mu\nu}=0$ everywhere except at the origin due to the mass pole.}. This fact has no consequences in the remaining discussions presented here, it simply is interpreted as that, for $\alpha_0>5$, in the neighborhood of the event horizon, the SC metric is a very good approximation to the background space-time.

Although the singularity at the event horizon is removed for constant exponent $\alpha_0\ge 3$, the space-time described by ELA metric~(\ref{ELA_metric}) has a singularity at the origin more
severe than the SC mass-pole. Specifically we obtain
\be
R_{\alpha_0\ge 3}(r_1\sim 0)\sim\frac{1}{r_1^{\alpha_0}}\ \ ,\ \ {\mathcal{R}}_{\alpha_0\ge 3}(r_1\sim 0)\sim\frac{1}{r_1^{2\alpha_0}}\ .
\ee
Hence, to strictly maintain the SC mass-pole at the origin we are left with the only
possible value for the constant exponent $\alpha_0=3$ for which space-time is not approximately
Ricci flat in the neighborhood of the event horizon $r_1=r_{1.\mathrm{SC}}$. We remark that the SC metric is usually used to describe space-time for most astrophysical systems and the estimative obtained agree to an high degree of accuracy with experimental measurements, hence we expect that near the massive central objects the properties of space-time coincide with the ones of an SC space-time, in particular that it is approximately Ricci flat within a neighborhood of the event horizon\footnote{The higher the value of $\alpha_0$, for a given accuracy, the greater the neighborhood for which space-time is approximately Ricci flat.}. If we wish to maintain this possibility it is enough
to consider a radial coordinate dependence of the exponent $\alpha$ of the form
\be
\alpha=\alpha_0+\alpha_1 U_{\mathrm{SC}}\ \ \ ,\ \ \alpha_1<0
\label{alpha}
\ee
where $\alpha_1$ is an arbitrarily small negative exponent which regularizes
the singularity at the origin being its effects, for most purposes, negligible
outside the event horizon $r_1>r_{1.\mathrm{SC}}$. This exponent can be justified
by noting that it correspond to the first order expansion in the
gravitational field of a generic spherically symmetric exponent function.

For this exponent we obtain that the curvature invariant at
the origin asymptotically coincides with the respective quantity for the Schwarzschild metric
independently of the value of the coefficient $\alpha_0$,
${\mathcal{R}}_{\alpha_0,\alpha_1<0}(r_1\sim 0)=R_{\mathrm{SC}}$~(\ref{SC_R}).
As for the SC radius we conclude that the event horizon is maintained for $\alpha_0+\alpha_1>1$ and  space-time has the following properties
\be
\ba{rcll}
\alpha_0+\alpha_1&\ge&3\ :& \mathrm{space-time\ is\ singularity\ free\ at\ }r_1=r_{1.\mathrm{SC}}\ ,\\[5mm]
\alpha_0+\alpha_1&>&5\ :& \mathrm{space-time\ is\ locally\ Ricci\ flat\ at\ }r_1=r_{1.\mathrm{SC}}\ .
\ea
\ee

\section{Energy-Momentum Density, Pressures and the Efective Equation of State}

It is missing to analyze the densities and pressures. In the following we will explicitly show that the pressures are anisotropic and that, for small $|\alpha_1|\ll 1$, the energy-momentum density is positive definite. Hence, neglecting the contributions of the radial dependence on the exponent outside the SC event horizon and using the notation $\bar{r}_1=c^2\,r_1/2GM$ for compactness of the equations, we obtain the following expressions for the energy density and pressures $\,_{(\alpha_0)}\rho_r$, $\,_{(\alpha_0)}p_r$, $\,_{(\alpha_0)}p_\theta=\,_{(\alpha_0)}p_\varphi$
\be
\ba{rcl}
\,_{(\alpha_0)}\rho&=&\displaystyle \frac{M}{2\pi\,r_1^2}\delta(r_1)+\frac{H^2}{8\pi\,G}\,Z^{\alpha_0-1}\left(3+\frac{\alpha_0-3}{\bar{r}_1}\right)\ ,\\
\,_{(\alpha_0)}p_{r}&=&\displaystyle\frac{c^2\,H^2}{8\pi\,G}\,Z^{\frac{\alpha_0}{2}-\frac{1}{2}}\left(2(1+q)-\frac{\alpha_0}{\bar{r}_1}Z^{\frac{\alpha_0}{2}-\frac{1}{2}}-3Z^{\frac{\alpha_0}{2}+\frac{1}{2}}\right)\ ,\\
\,_{(\alpha_0)}p_{\theta}&=&\displaystyle-\frac{c^2\,H^2}{8\pi\,G}\,Z^{\alpha_0-2}\left(3+\frac{2(\alpha_0-3)}{\bar{r}_1}+\frac{\alpha_0(\alpha_0-5)+6}{2\bar{r}_1^2}\right)\\
&&\displaystyle\hfill+\frac{c^2\,H^2}{8\pi\,G}\,(1+q)Z^{\frac{\alpha_0}{2}-\frac{3}{2}}\left(2+\frac{\alpha_0-4}{2\bar{r}_1}\right)\\
\ea
\label{rho_p_gen_r}
\ee
These expressions are plotted in figure~\ref{fig.rho} for distinct values of $\alpha_0$.
\fig{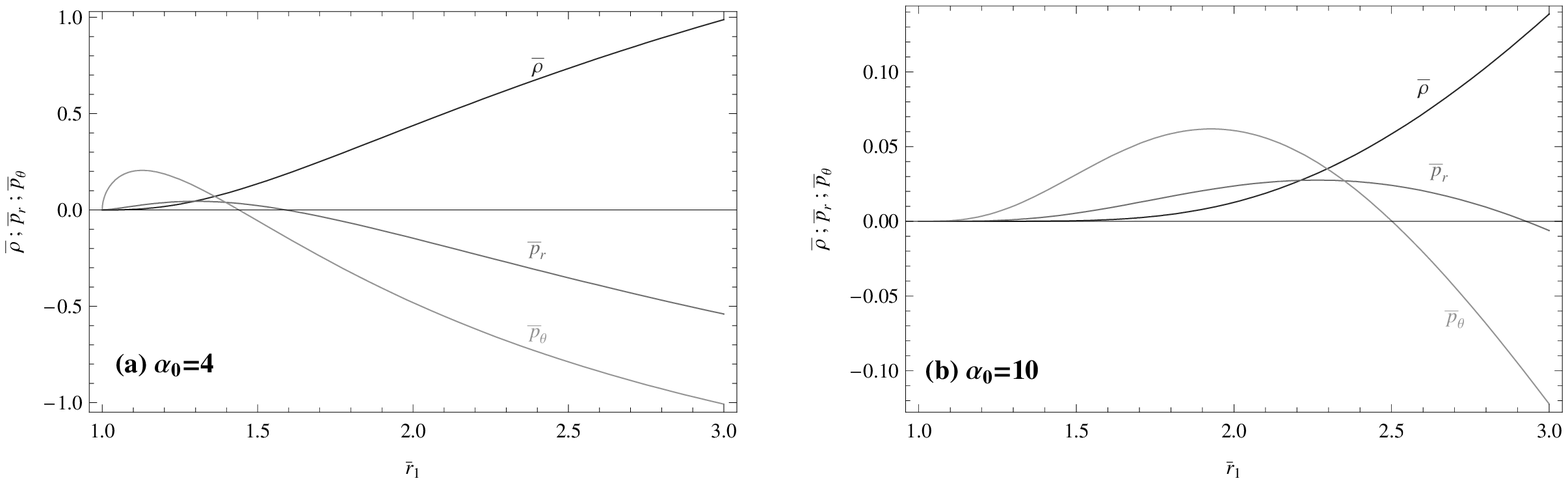}{130mm}{Scaled energy-momentum density $\bar{\rho}=(8\pi G/H^2)_{(\alpha_0)}\rho$ and anisotropic pressures $\bar{p}_r=(8\pi G/(cH)^2)_{(\alpha_0)}p_r$ and $\bar{p}_\theta=(8\pi G/(cH)^2)_{(\alpha_0)}p_\theta$ for the ELA metric~(\ref{ELA_metric}) with constant exponent {\textbf{(a)}} $\alpha_0=4$ and {\textbf{(b)}} $\alpha_0=10$.}{fig.rho}

As expected, at the origin, these quantities coincide with the respective ones for the SC metric and, at spatial infinity, with the ones for the RW metric. In addition we straight forward conclude
that, although spherical symmetry is maintained ($\,_{(\alpha_0)}p_{\theta}=\,_{(\alpha_0)}p_{\varphi}$), space-time is, generally, anisotropic ($\,_{(\alpha_0)}p_{\theta}\neq\,_{(\alpha_0)}p_{r}$) being only asymptotically isotropic at spatial infinity. Furthermore, for $\alpha_0\ge 3$, outside the event horizon,
the energy-momentum density is positive definite, being null only at the event horizon. This is a good feature of the metric since it allows a standard matter interpretation with the above density distribution and a description in terms of an effective matter density. We also note that for $\alpha_0>5$ all these quantities and respective derivatives are null at the event horizon $r_1=r_{1.\mathrm{SC}}$, $\,_{(\alpha_0)}\rho=\,_{(\alpha_0)}p_{r}=\,_{(\alpha_0)}p_{\theta}=\,_{(\alpha_0)}\rho'=\,_{(\alpha_0)}p'_{r}=\,_{(\alpha_0)}p'_{\theta}=0$
coinciding with the respective SC metric quantities and consistently with space-time being approximately Ricci flat near the event horizon.

As for the anisotropic equations of state $\,_{(\alpha_0)}\omega_r=\,_{(\alpha_0)}p_r/(c^2\,_{(\alpha_0)}\rho)$ and $\,_{(\alpha_0)}\omega_\theta=\,_{(\alpha_0)}\omega_\varphi=\,_{(\alpha_0)}p_\theta/(c^2\,_{(\alpha_0)}\rho)$
are plotted in figure~\ref{fig.w} for several values of $\alpha_0$.
\fig{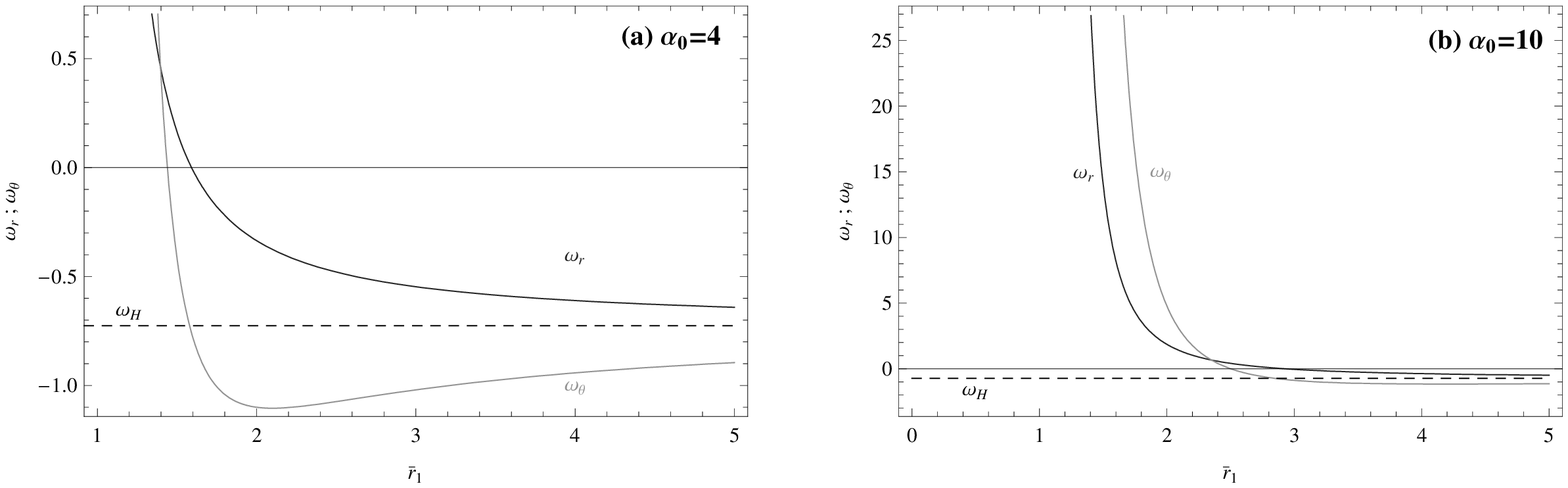}{130mm}{The equation of state parameters $\,_{(\alpha_0)}\omega_r$ and $\,_{(\alpha_0)}\omega_\theta$ for the ELA metric~(\ref{ELA_metric}) and $\omega_H$ for the RW metric~(\ref{RW_metric}) with constant exponent {\textbf{(a)}} $\alpha_0=4$ and {\textbf{(b)}} $\alpha_0=10$.}{fig.w}

Both the parameters of the equation of state, $\,_{(\alpha_0)}\omega_r$ and $\,_{(\alpha_0)}\omega_\theta$, diverge at the event horizon (this is due to the existence of the horizon itself) being positive near this horizon, as it is characteristic of standard attractive matter, and as we move away for larger values of $r_1$, become negative due to the expansion effects. At spatial infinity both parameters converge to the RW metric equation of state parameter, $\omega_H=-(1-2q)/3$. These results are consistent with the interpretation of the ELA metric~(\ref{ELA_metric}) being an interpolation between the point-like matter distributions and the cosmological background, near the massive objects standard matter interaction dominates while for large radii the repulsive effects (mostly due to the cosmological constant) dominate. However we note that a direct relation with the SC metric cannot be drawn in this discussion. This metric has $p_{\mathrm{SC}}=\rho_{\mathrm{SC}}=0$ everywhere except at the origin and an equation of state cannot be defined.

\section{Conclusions}

In this letter we have presented a short derivation and analysis of an anisotropic
metric~(\ref{ELA_metric}) with a radial coordinate exponent $\alpha$~(\ref{alpha})
describing point-like matter in an expanding background, hence
interpolating between the SC metric~(\ref{SC_metric}) and the RW metric~(\ref{RW_metric}).
For $\alpha_0+\alpha_1\ge 3$ with $-1\ll\alpha_1<0$, the SC event horizon is mantained being
non-singular and the singularity at the origin coincides with the SC mass-pole. Furthermore the
energy-momentum density is positive definite outside of the event horizon allowing for a effective
standard matter density interpretation. If it turns out that the ELA metric~(\ref{ELA_metric})
correctly describes the physical reality for specific values of the coefficients $\alpha_0$ and
$\alpha_1$ it may solve the long standing puzzle originally addressed by McVittie~[\citenum{McVittie}].
If possible to directly test this metric it will be, most probably, through its corrections to the
gravitational attraction law for space-times described by the SC metric. We will address this issue
somewhere else~[\citenum{e-print}].

\noindent{\bf Acknowledgements}-- The author acknowledges and thanks the referee for comments and suggestions that significantly improved the manuscript. This work was supported by grant SFRH/BPD/34566/2007 from FCT-MCTES.

\end{document}